# Enlarged magnetic focusing radius of photoinduced ballistic currents

*Markus Stallhofer,[1] Christoph Kastl,[1] Marcel Brändlein,[1] Dieter Schuh,[2] Werner Wegscheider,[3] Jörg P. Kotthaus,[4] Gerhard Abstreiter,[1] and Alexander W. Holleitner[1*]*

1 Walter Schottky Institut and Physik-Department, Technische Universität München, Am Coulombwall 4a, D-85748 Garching, Germany

2 Institut für Experimentelle und Angewandte Physik, Universität Regensburg, D-93040 Regensburg, Germany

3 Laboratorium für Festkörperphysik, HPF E 7, ETH Zürich, 8093 Zürich, Switzerland

4 Ludwig-Maximilians Universität München and Center for Nanoscience, Geschwister-Scholl-Platz 1, D-80539 München, Germany

Corresponding author: Alexander W. Holleitner; e-mail: holleitner@wsi.tum.de

ABSTRACT: We exploit GaAs-based quantum point contacts as mesoscopic detectors to analyze the ballistic flow of photogenerated electrons in a two-dimensional electron gas at a perpendicular magnetic field. Whereas charge transport experiments always measure the classical cyclotron radius, we show that this changes dramatically when detecting the photoinduced non-equilibrium current in magnetic fields. The experimentally determined radius of the trajectories surprisingly exceeds the classical cyclotron value by far. Monte Carlo simulations suggest electron-electron scattering as the underlying reason.

In the presence of a magnetic field $\boldsymbol{B}$, propagating charge carriers obey the Lorenz force and move on curved trajectories in the plane perpendicular to the field. The corresponding classical cyclotron radius $r_{\text{cyclo}}(|\boldsymbol{B}|) = \frac{\boldsymbol{m}|\boldsymbol{v}|}{q\,|\boldsymbol{B}|}$ (1) for a charge $q$ is a measure of its momentum and thus contains both mass $m$ and velocity $\boldsymbol{v}$ of the carrier. Based on Eq. 1, cyclotron resonance experiments on two-dimensional electron systems have been used to verify the 2D nature, to determine the components of the effective mass tensor in solids and to obtain information on Fermi contours [1]. However, following Kohn's theorem, the mass measured in cyclotron resonance experiments is expected to be the bare mass, not affected by electron-electron interactions [2]. In magnetic focusing experiments, quantum point contacts (QPCs) have been used both as injectors and detectors to indirectly obtain information on cyclotron trajectories of ballistic currents in two-dimensional electron gases (2DEGs) [3,4,5,6]. At 2DEGs' boundaries, even skipping cyclotron orbits have been visualized using a scanning gate technique [7,8,9,10], and different electron trajectories have been shown to interfere with each other [8,11,12]. The trajectories extracted were in accordance with Eq. (1) and the expected effective mass $m^{*}_{\text{GaAs}}$ of the GaAs-based circuits.

Here, we exploit an optical beam induced current (OBIC) technique to directly analyze the flow-patterns of photogenerated electrons in a 2DEG embedded in an AlGaAs/GaAs quantum well at a perpendicular magnetic field. In contrast to the magnetic focusing results gained from transport experiments, we uncover the magnetic focusing dynamics of photogenerated non-equilibrium excess charge carriers in a 2DEG. A laser locally creates charge carriers at a certain position in the 2DEG, and the ballistic photocurrent of the photogenerated electrons across an adjacent QPC is measured as a function of the laser position [13,14]. This OBIC-technique allows adjusting the excitation position independent of the applied $|\boldsymbol{B}|$. Thus, it is possible to directly map the magnetic flow patterns of photogenerated electrons in the 2DEG. We observe curved trajectories with a radius being inversely proportional to $|\boldsymbol{B}|$ as expected from Eq. (1). However, for both resonant and non-resonant optical excitation of the quantum well, the radii of the measured trajectories are 10 to 30 times larger than anticipated. Peculiarly, assuming Eq. (1) and the given electron Fermi-velocity $|\boldsymbol{v}_{\text{Fermi}}|$ in the 2DEG, a factor of 30 would suggest an electron mass for the GaAs quantum well which is up to two times larger than the vacuum value $m_e$, with $30 \cdot m^*_{GaAs} = 30 \cdot 0.067 \cdot m_e \approx 2 \cdot m_e$. We can exclude plasmons to be the origin of the observations. Instead, we explain the findings by electron-electron scattering processes [15,16,17]. We perform Monte Carlo simulations of the effective radius of photogenerated excess electrons in a perpendicular magnetic field at the presence of small-angle scatterers. We find very good agreement between the simulations and the data. Our findings suggest that due to an enhanced influence of electron-electron scattering, the radius of the trajectories of hot ballistic electrons is generally enlarged as compared to Eq. (1). Our observations underline the predominant influence of electron-electron scattering processes in mesoscopic and nanoscale photodetectors.

Starting point is an AlGaAs/GaAs-heterostructure with a 25 nm wide GaAs quantum well 95 nm below the surface. The quantum well comprises a 2DEG with a Fermi energy $E_{Fermi}$ = 9.83 meV and an electron mobility $\mu = 1.74 \times 10^6$ $cm^2$/Vs. The mean free path $l_{mfp}$ is determined to be 15.1 µm at $T$ = 4.2 K. Generally, $l_{mfp}$ is the averaged length scale, on which majority charge carriers can propagate without momentum scattering. In GaAs-based heterostructures at low temperatures, $l_{mfp}$ is mostly limited by long-range small-angle scatterers. As sketched in Fig. 1(a), QPCs are lithographically defined by shallow-etching two circles with radius of $r_{circuit}$ = 15 µm defining an electronic constriction [13]. The lithographic width of the constriction is ~300 nm. Considering depletion, the value translates to an electronic width of ~50 nm, which is comparable to the Fermi wavelength $\lambda_{Fermi}$ = 48 nm. On both ends of this QPC, the remaining 2DEG acts as source and drain. The QPC is covered with an opaque gold topgate with a lateral width of 8 µm, and the topgate is connected to a voltage $V_G$. Without laser excitation, the low-temperature conductance across the QPC shows quantization steps of $2e^2/h$ [3,4,19]. The OBIC measurements are performed at photon energies $E_{ph,1}$ = 1.546 eV and $E_{ph,2}$ = 1.598 eV. Under laser excitation, the electron temperature is estimated to be 8-10 K [14]. $E_{ph,1}$ corresponds to a resonant excitation of the quantum well, as proven by independent photoluminescence measurements [data not shown]. For $E_{ph,2}$, the photogenerated electrons exhibit an average excess energy of $\Delta E$ = 49 meV. The energetic FWHM of the laser is 6.4 meV. Photogenerated electrons in the source contact propagate ballistically across the QPC before they are detected in drain. We note that we can neglect the effect of the photogenerated holes in the presented experiment, because the photocurrent read-out is triggered by the laser excitation in the MHz-regime as thoroughly discussed in refs. [13,14]. For a fixed excitation position with respect to the QPC, the amplitude of the OBIC shows clear quantization steps as a function of $V_G$ [filled triangles in Fig. 1(b)]. The steps are interpreted to reflect the electronic 1D subbands of the QPC. The peak at the onset of the first plateau [open triangle in Fig. 1(b)] can be explained by an interference effect [20] or the influence of an impurity at the aperture of the QPC [21,22], and it is beyond the scope of the present manuscript.

By simultaneously detecting the OBIC and the reflected light from the sample [Fig. 1(c)], one can relate the spatial dependence of the photocurrent to the sample geometry. Fig. 1(d) depicts a respective OBIC map at $|\boldsymbol{B}| = 0$ T. The OBIC decays on a typical decay length $l_{\text{photo}}$ comparable to $l_{\text{mfp}}$ [13]. At a moderate magnetic field, the OBIC signal is not symmetric anymore [Figs. 1(e) and (f)]. Experimentally, we find that for $|\boldsymbol{B}| \geq 100$ mT, the minimum detectable radius of the trajectories is limited by $r_{\text{circuit}}$. This can be understood such that the photogenerated electrons propagate in skipping orbits smaller than the laser spot size of ~2 µm along the circuit's boundary. In the following, we focus on magnetic fields with $|\boldsymbol{B}| < 100$ mT, for which photogenerated electrons should propagate along cyclotron orbits with $r_{\text{cyclo}} > 0.87$ µm according to Eq. (1).

Such an OBIC map is depicted as dots in Fig. 2(a). The QPC is positioned at $(x, y) = (0, 0)$, and the topgate extends from -4 µm $\leq x \leq$ 4 µm. The data are fitted by a 2D function $h(x, y)$:

$$h(x,y) \coloneqq e^{\frac{-\left(\sqrt{(x-x_C)^2+(y-y_C)^2}-r_{\text{eff}}\right)^2}{2\left(\sigma_0\cdot\left[1+\left(\frac{r_{\text{eff}}\cdot\Phi}{a}\right)^b\right]\right)^2}} \cdot \Theta(\Phi) \cdot e^{-\frac{\Phi\cdot r_{\text{eff}}}{l_{\text{photo}}}} * \frac{1}{\sqrt{2\pi}\cdot\sigma_C} e^{\frac{-(\Phi\cdot r_{\text{eff}})^2}{2\sigma_C^2}}, \qquad (2)$$

with radius $r_{\text{eff}}$ and an angle $\phi$ denoting the polar coordinates of the circular trajectories around a center point $(x_c, y_c)$ (see [19] for details). $\sigma_0$ represents the spatial width of the Gaussian distribution of the photogenerated electrons at the aperture of the QPC at ($x$ = 5 µm, $y$ = 0). The width widens along the trajectories as described by the parameters $a$ and $b$. The spatial FWHM of the laser spot is described by $\sigma_C$. In order to demonstrate the very good agreement of $h(x,y)$ with the data, we depict

Cartesian cuts along $y$ at $x$ = 6 µm [Fig. 2(b)] and along $x$ at $y$ = 0 µm [Fig. 2(c)]. The white line in Fig. 2(a) highlights the fitted circular orbit of the photogenerated electrons along the magnetic focusing trajectory. Such traces for varying $\boldsymbol{B}$ are plotted in Figs. 2(d) and (e) for the QPC tuned to the first and second plateau, respectively. Surprisingly, the values for $r_{\mathrm{eff}}$ are approximately one order of magnitude larger than the classical cyclotron radius $r_{\mathrm{cyclo}}$. For comparison, for $|\boldsymbol{B}|$ = 50 mT Eq.(1) gives $r_{\mathrm{cyclo}}$ = 1.74 µm, whereas we measure $r_{\mathrm{eff}}$ = 32 µm for $E_{\mathrm{ph}}$ = 1.546eV and $P_{\mathrm{laser}}$ = 2.8 µW. For the obvious offset in $|\boldsymbol{B}|$ see [19].

Fig. (3) shows the inverse of $r_{\mathrm{eff}}$ vs. $\boldsymbol{B}$ for resonant excitation [Fig. 3 (a) and Fig. 3 (b)] and non-resonant excitation [Fig. 3 (c) and Fig. 3 (d)] for different $P_{\mathrm{Laser}}$. The data are shown as scattered points. The lines are fits according to Eq. (1) with $b = m^*|\boldsymbol{v}|/e$ as a fitting parameter. Generally, assuming a ballistic cyclotron motion of single electrons after a resonant excitation at $E_{\mathrm{Fermi}}$ = 9.83 meV, $b$ can be estimated to be $b_{\mathrm{Fermi}} = m^*|\boldsymbol{v}_{\mathrm{Fermi}}|/e$ = 87 µm · mT. For a non-resonant excitation with an excess energy of $\Delta E$ = 49 meV, one expects $b$ to be 192 µm · mT ~ 2.2 × $b_{\mathrm{Fermi}}$ [19]. The experimentally determined values of $b$ are depicted in Fig. 3(e). Surprisingly, they are 10 to 30 times larger than $b_{\mathrm{Fermi}}$. Additionally, $b$ does not significantly depend on $E_{\mathrm{Photon}}$ nor does it differ for the QPC tuned to the 1$^{\mathrm{st}}$ or 2$^{\mathrm{nd}}$ conductance plateau. However, $b$ increases with $P_{\mathrm{Laser}}$ and thus, with the density of the photogenerated electrons. In turn, electron-electron scattering most probably deflects the cyclotron traces of individual photogenerated electrons yielding an enlarged effective magnetic focusing radius, as discussed in the following.

The average transit time $\tau_{\mathrm{transit}}$ of a resonantly excited electron from the laser spot to the QPC can be estimated to be $\tau_{\mathrm{transit}} \approx l_{\mathrm{photo}}/|\boldsymbol{v}_{\mathrm{Fermi}}| \approx l_{\mathrm{mfp}}/ |\boldsymbol{v}_{\mathrm{Fermi}}|$ = 15.1 $\mu$m/ 2.3 × 10$^5$ ms$^{-1}$ ≈ 66 ps. At an excess energy $\varDelta E$ of only a few meV above $E_{\mathrm{Fermi}}$, the electron trajectories in 2DEGs are dominated by electron-electron scattering [15,16]. The electron-electron scattering time in 2DEGs is inversely proportional to $\varDelta E^2 \ln \varDelta E$, and for $\varDelta E \sim$ meV, it is in the range of a few ps [23]. This relates to an unperturbed motion along only a few hundreds of nanometer, which is much shorter than $l_{\mathrm{mfp}}$ and $r_{\mathrm{eff}}$ presented here. Hence, photogenerated excess electrons can be assumed to scatter at a few electrons during $\tau_{\mathrm{transit}}$ before they reach the QPC. We note that the energy distribution of photogenerated electrons approaching the QPC can be estimated to be a combination of a Gaussian and a Maxwell-Boltzmann distribution [24,25]. Since we detect quantized photocurrent steps [Fig. 1(b)], an upper limit of the FWHM of this effective energy distribution can be estimated to be smaller than the 1D subband spacing of the QPC of ~4 meV [13,19] for both resonant and non-resonant excitation. At $\Delta E$ = 49 meV (non-resonant excitation), the photogenerated electrons strongly interact with longitudinal optical phonons. Hereby, they effectively relax near to the Fermi edge within a few picoseconds [26]. In combination with the enhanced electron-electron scattering for non-equilibrium charge carriers, this substantially explains why there is no significant variation between the two photon energies.

Based upon these arguments, we perform Monte Carlo simulations to describe the enlarged magnetic focusing radii by the following simplified model. An electron resonantly created at the laser spot propagates at $|\boldsymbol{v}_{\mathrm{Fermi}}|$, and it obeys the Lorentz force. The propagation length of the electron is exponentially weighted by the mean free path in order to consider momentum scattering. On top, the electron is scattered at other electrons with a rate $\tau_{\mathrm{ee}}^{-1}$ on its way to the QPC [19]. Each electron-electron scattering event deflects the $k$-vector of the electron at a certain small angle, as recently discussed [27,28]. In addition, the boundary of the 2DEG allows skipping orbits with a certain loss probability of $p_{\mathrm{spec}}$. Fig. 4(a) exemplifies two of such trajectories which eventually reach the QPC. For each excitation position $(x,y)$, only a certain number of all possible trajectories reach the QPC, since the laser excitation creates electrons isotropically and the trajectories are weighted by the mean free path. This defines a hit rate $p_{\mathrm{hit}} = p_{\mathrm{hit}}(x,y,\boldsymbol{B})$, which is exemplarily plotted for certain simulation parameters in Fig. 4(b). Generally, we find that the simulations nicely reproduce the experimental data as in Fig. 2(a). In a next step, the simulated data are fitted with $h$(x,y) and an effective magnetic focusing radius $r_{\mathrm{eff}}^{\mathrm{sim}}$, analogously to the fitting procedure of the experimentally determined data. Again, we observe that $r_{\mathrm{eff}}^{\mathrm{sim}}$ is inversely proportional to $|\boldsymbol{B}|$ [Fig. 4(c)]. On the one hand, this finding confirms that the Monte Carlo simulation only considers elastic electron-electron scattering events in combination with the Lorentz force. On the other hand, the slopes in Fig. 4(c) again depend on $\tau_{\mathrm{ee}}^{-1}$. In other words, $r_{\mathrm{eff}}^{\mathrm{sim}}$ reflects the number of scattering events along the trajectories. A detailed evaluation of this dependence is shown in Fig. 4(d). Most importantly, the simulated values $b_{\text{sim}}$ are up to 50 times larger than the anticipated $b_{\text{Fermi}}$. For typical electron-electron scattering rates $1\frac{1}{\text{ps}} \leq \frac{1}{\tau_{\text{ee}}} \leq 3\frac{1}{\text{ps}}$ [25], $b_{\text{sim}}$ agrees well with the experimentally determined values in Fig. 3(e), although the simulation does not contain any scaling parameter.

Strictly speaking, the assumed scattering mechanism is a generalized small-angle scattering including electron-electron scattering. However, the underlying scattering process relates to a length scale of typically ~100 nm, which is significantly shorter than $l_{\text{mfp}}$. In other words, the scattering rate $\frac{1}{\tau_{\text{ee}}}$ does not correspond to the long-range small angle scattering processes which influence $l_{\text{mfp}}$. Instead, the electron-electron scattering is the predominant scattering mechanism for the photoinduced non-equilibrium charge carriers. This interpretation is consistent with the fact that there is no significant variation between the two photon energies [Fig. 3], which vary by more than the excitation energy of an optical phonon. We further note that momentum conservation is not necessarily fulfilled, since not all electrons initially created by the laser position are detected at the QPC. Photogenerated electrons are scattered into all directions. In our simulations, momentum conservation is therefore not assumed.

In principle, the large value of $r_{\text{eff}}$ could also point towards collective phenomena such as plasmons or charge density waves [29,30]. Both can influence the screening of the lattice, affecting $m^*$, and the screening of scatterers, as has been reported for low temperature grown GaAs [29,31]. Collective phenomena, however, cannot account for an increase of 30 times. In addition, we measure a very similar increase of $r_{\text{eff}}$ for both resonant and non-resonant excitation. Only for the non-resonant excitation, the Fermi-gas would have enough excess energy to launch plasmons. We furthermore can exclude streaming motion to be the underlying reason for our experimental findings [32]. For streaming motion, one would expect straight propagation paths for the electrons whereas we clearly observe curved trajectories. This is consistent with the fact that streaming motion occurs only in crossed electric and magnetic fields, which can be assumed to be not the case at the positions of the laser excitation [Fig. 1(c) - (f)]. Furthermore, assuming a macroscopic (diffusive) Hall-circuit with crossed electric and magnetic fields, the classical Hall angle reads $\theta_{\text{Hall}} = \arctan(\mu|\boldsymbol{B}|)$. For the given $\mu$ and a magnetic field of $|\boldsymbol{B}_z| = 10$ mT, $\theta_{Hall}$ would already be 60°. Generally, in a notional diffusive regime, the charge carriers are deflected along straight lines with a direction given by $\theta_{Hall}$, which we do not detect in our OBIC-maps [Fig. 1(e) and (f)]

We further note that the data in Fig. 3(e) suggest an enlarged $r_{eff}$ even at $P_{laser} = 0$. This is consistent with the fact that electron-electron scattering is present in the 2DEG without laser excitation [23,25]. In contrast to previous magnetic focusing experiments [5,6,11], however, our measurements exclusively concentrate on the propagation paths of photogenerated, non-equilibrium excess electrons. They have a softened energy distribution compared to the one of the thermalized, non-excited electrons [24]. Our results demonstrate that photogenerated electrons experience an increased electron-electron scattering compared to thermalized non-excited electrons, which is a unique fingerprint of optoelectronic experiments.

In summary, we analyze the magnetic focusing of photogenerated excess charge carriers in a GaAs 2DEG. By applying a photocurrent mapping technique, we spatially resolve the trajectories of photogenerated electrons in a perpendicular magnetic field. Initially assuming, that the electronic motion can be classically described with cyclotron orbits, we model the data with a two-dimensional fitting function that enables us to extract the classical cyclotron radius as a fitting parameter. Surprisingly, we find that the measured values lie significantly above the expected values. By performing Monte-Carlo simulations we attribute the enlarged magnetic focusing radius to electron-electron scattering.

ACKNOWLEDGMENT. We thank A. Högele for technical support and G. J. Schinner, A. Govorov, and L. Prechtel for fruitful discussions. Financial support from the German Science Foundation DFG (Ho 3324/4), the Center for NanoScience (CeNS), and the German excellence initiative via the “Nanosystems Initiative Munich (NIM)” is gratefully acknowledged. M. Stallhofer gratefully acknowledges the support of the TUM Graduate School’s Faculty Graduate Center of Physics at the Technische Universität München.

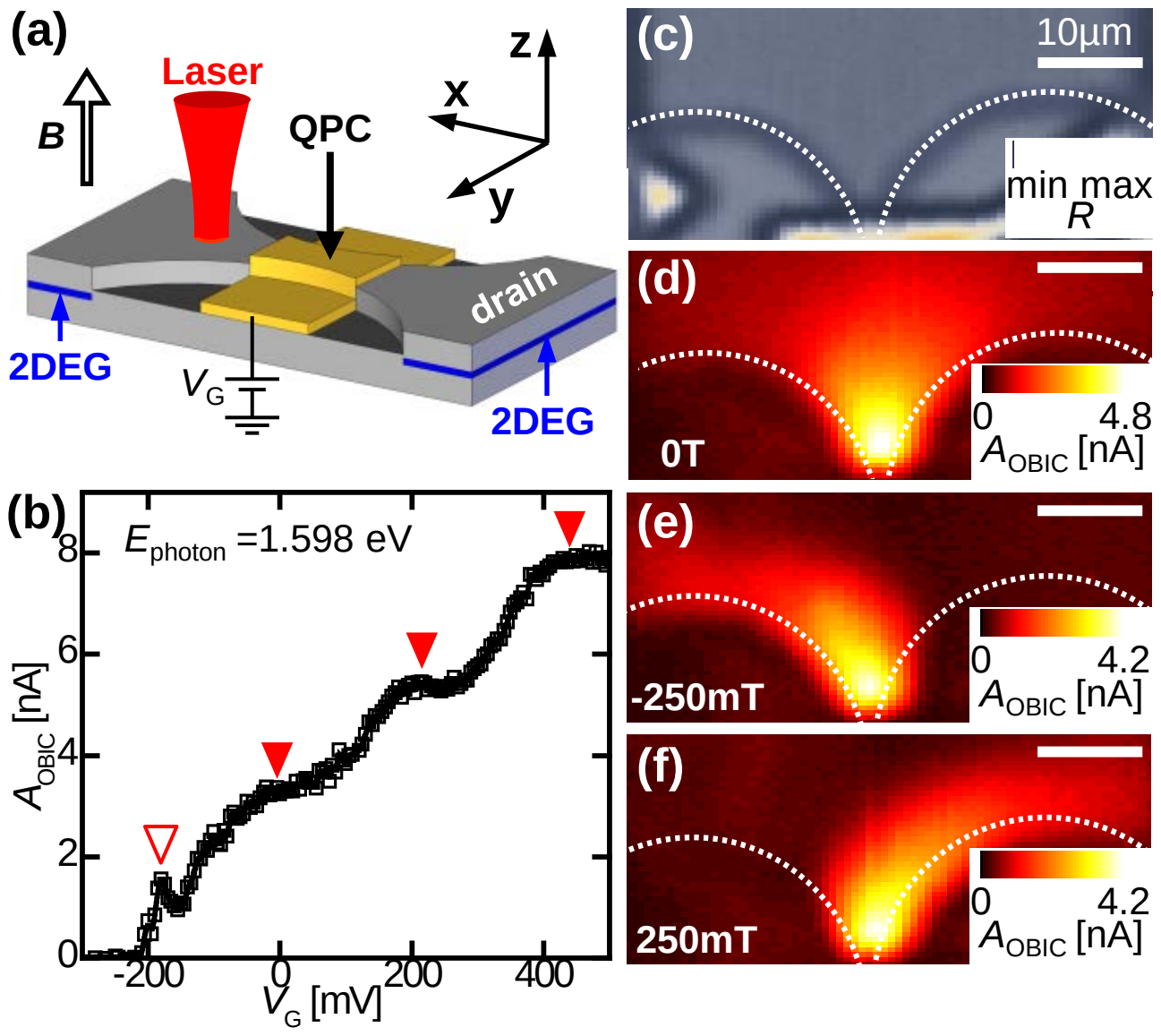

FIG. 1 (color online). (a) Sketch of a quantum point contact (QPC) with an opaque topgate, a laser excitation in the 2DEG section acting as source, and a perpendicularly applied magnetic field ***B***. The photocurrent is detected at drain. (b) $A_{OBIC}$ as function of $V_G$. Quantization steps reveal one-dimensional subbands (filled triangles) ($V_{SD}$ = -2.5 mV, $E_{ph}$ = 1.598 eV, $P_{laser}$ = 0.54 µW, |***B***| = 0 T). (c) Spatially resolved reflection map. (d)-(f) Spatial map of $A_{OBIC}$ for (d) 0 T, (e) -250 mT, and (f) 250 mT ($V_{SD}$ = -2.5 mV, $V_G$ = 78 mV, $E_{ph}$ = 1.546 eV, $P_{laser}$ = 1.1 µW).

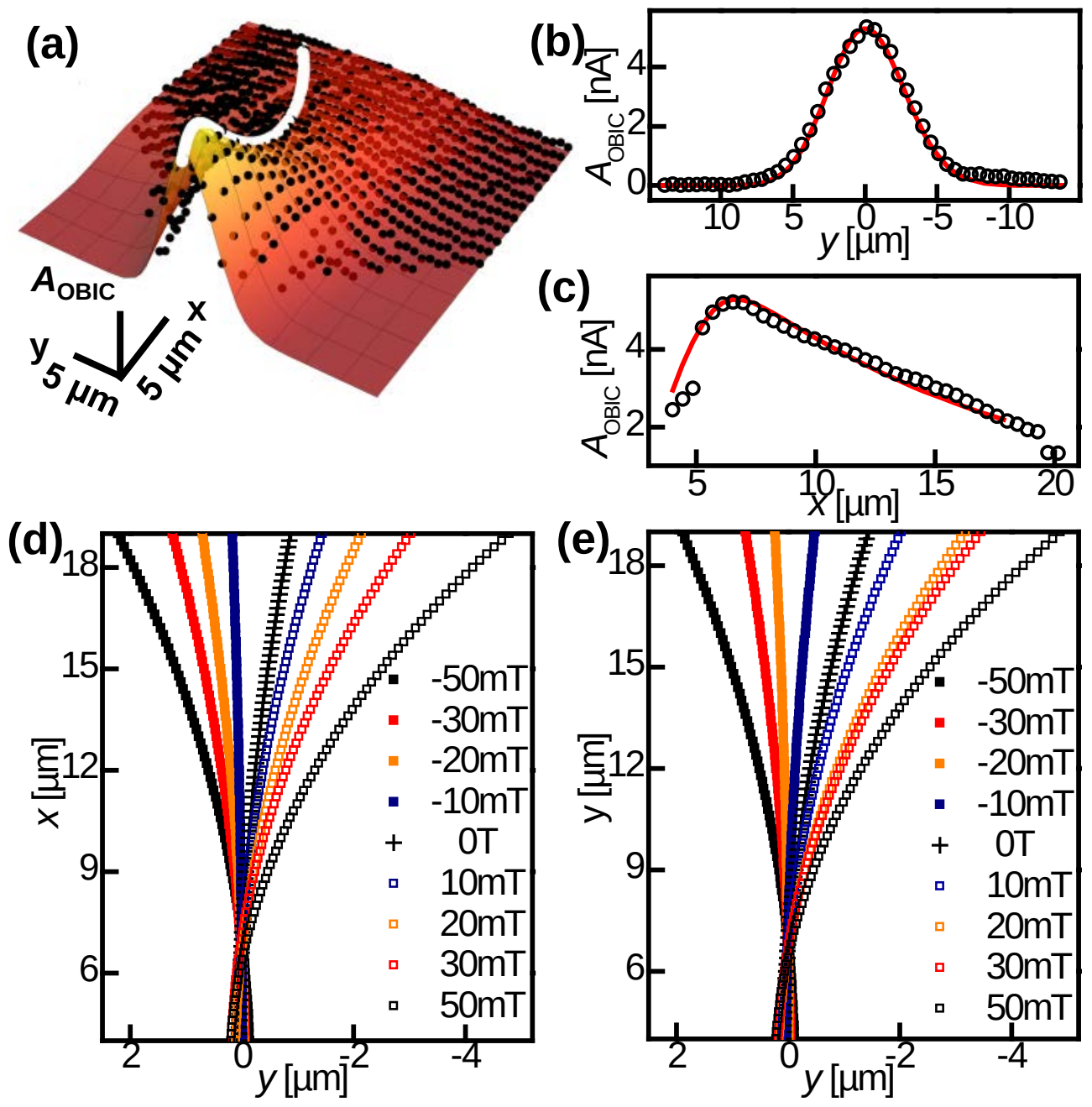

FIG. 2 (color online). (a) $A_{\mathrm{OBIC}}$ map for resonant excitation (scattered black points) with two-dimensional fitting function ($E_{\mathrm{ph}}$ = 1.546 eV, $V_{\mathrm{G}}$ = 0 mV, $V_{\mathrm{SD}}$ = -2.5 mV, $P_{\mathrm{laser}}$ = 1.0 µW, $\boldsymbol{B}$ = -50 mT). White line denotes cyclotron trace of maximum $A_{\mathrm{OBIC}}$. (b) and (c): Cuts along *x* and *y* in Fig. 2(a), with scattered data points and fitting function ($V_{\mathrm{G}}$ = 60 mV, $V_{\mathrm{SD}}$ = -2.5 mV, $P_{\mathrm{laser}}$ = 1.1 µW, $\boldsymbol{B}$ = 50 mT). (d) and (e): Cyclotron trajectories extracted from OBIC maps for the 1$^{\mathrm{st}}$ ($V_{\mathrm{G}}$ = 60 mV) and 2$^{\mathrm{nd}}$ ($V_{\mathrm{G}}$ = 250 mV) quantization plateaus, respectively ($V_{\mathrm{SD}}$ = -2.5 mV, $P_{\mathrm{laser}}$ = 1.1 µW, $E_{\mathrm{ph}}$ = 1.598 eV).

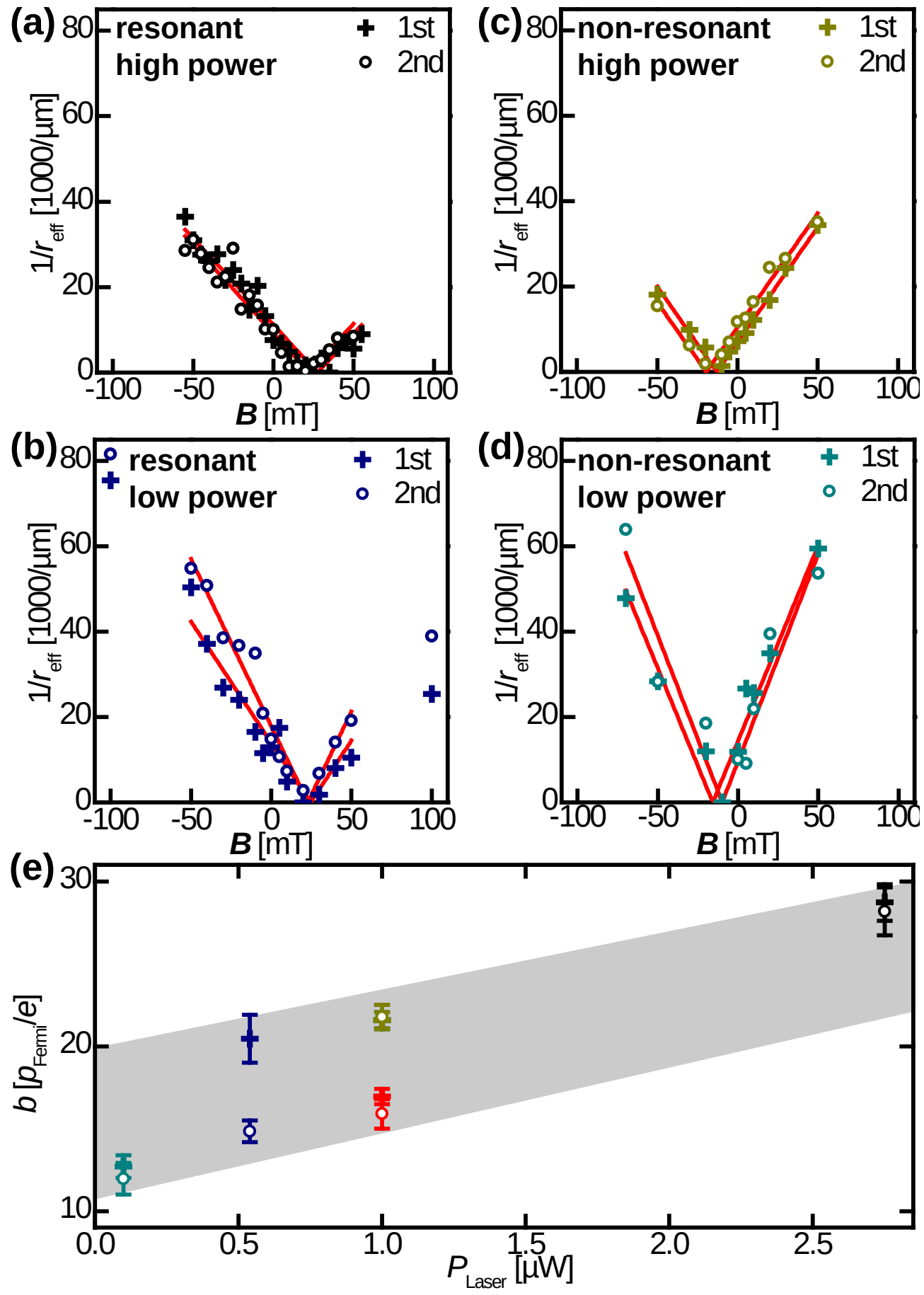

FIG. 3 (color online). Inverse $r_{eff}$ vs. $\boldsymbol{B}$ for the 1st (crosses) and 2nd (circles) quantization plateau at (a) $E_{ph}$ = 1.546 eV and $P_{laser}$ = 2.8 µW, (b) $E_{ph}$ = 1.546 eV and $P_{laser}$ = 0.54 µW, (c) $E_{ph}$ = 1.598 eV and $P_{laser}$ = 1.1 µW, and (d) $E_{ph}$ = 1.598 eV and $P_{laser}$ = 0.11 µW. Lines are fits. (e) Fitting parameter $b$ extracted from Figs. 3(a) – (d) as a function of $P_{laser}$. Red symbols represent data for resonant excitation $E_{ph}$ = 1.546 eV at $P_{laser}$ = 1.0 µW.

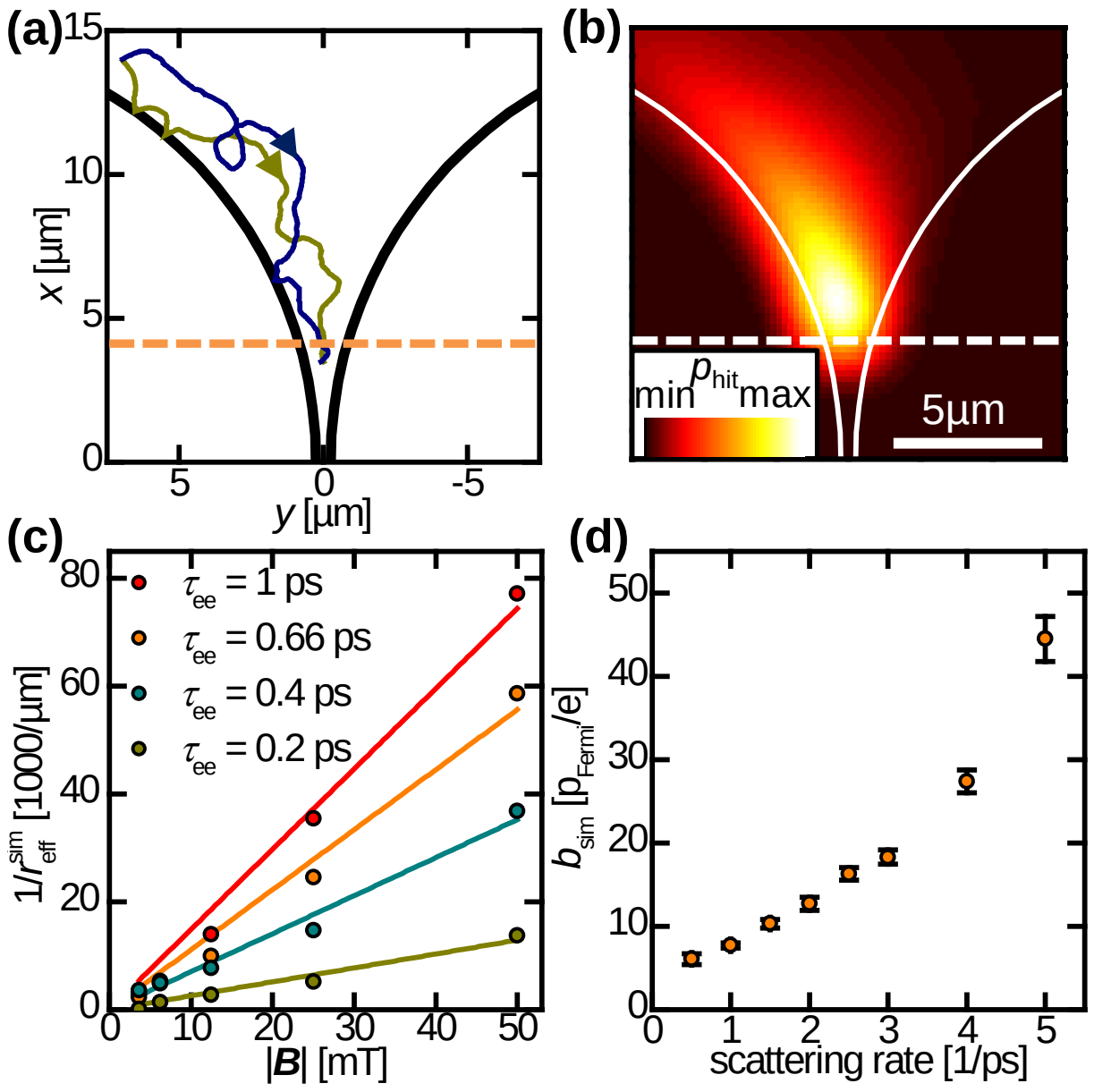

FIG. 4 (color online). (a) Typical simulated trajectories with a common starting point. Black lines represent circuit boundaries. Dashed line indicates boundary of topgate. (b) Simulated spatial distribution of $p_{hit}$ for $|\boldsymbol{B}| = 50$ mT and $\tau_{ee} = 2$ ps. (c) Extracted inverse of $r_{eff}^{sim}$ vs. $|\boldsymbol{B}|$ for varying $\tau_{ee}$. (d) Simulated inverse slopes $b_{sim}$ vs. scattering rate.

REFERENCES.

# Supplementary information:

# Enlarged magnetic focusing radius of photoinduced ballistic currents

*Markus Stallhofer,*[1] *Christoph Kastl,*[1] *Marcel Brändlein,*[1] *Dieter Schuh,*[2] *Werner Wegscheider,*[3] *Jörg P. Kotthaus*[4]*, Gerhard Abstreiter*[1]*, and Alexander W. Holleitner*[1*]

1 Walter Schottky Institut and Physik-Department, Technische Universität München, Am Coulombwall 4a, D-85748 Garching, Germany

2 Institut für Experimentelle und Angewandte Physik, Universität Regensburg, D-93040 Regensburg, Germany

3 Laboratorium für Festkörperphysik, HPF E 7, ETH Zürich, Schafmattstr. 16, 8093 Zürich, Switzerland

4 Ludwig-Maximilians Universität München and Center for Nanoscience, Geschwister-Scholl-Platz 1, D-80539 München, Germany

Corresponding author: Alexander W. Holleitner; e-mail: holleitner@wsi.tum.de

## 1. Experimental details

The sample is placed inside a confocal microscope in a Helium bath cryostat at temperature $T$ = 4.2 K. Due to laser and current heating as well as heat radiation we assume an electron temperature of 8 - 10 K. Supplementary Fig. S1 depicts conductance curves through the QPC after illumination [Fig. S1(a)] and during illumination [Fig. S1(b)]. The conductance quantization in units of $2e^2/h$ is clearly visible even at the relatively high electron temperature.

The position of the sample with respect to the laser is controlled by a resistive element on the nanopositioners (attocube ANPx101/Res) with a relative precision of approx. 200 nm. Between different measurements the repeatability precision of the absolute position is approximately 1 µm. While scanning the focused laser spot across the sample, we measure the optical beam induced current (OBIC) through the QPC depending on the excitation position with the current voltage amplifier being connected to a fast lock-in amplifier that is triggered to the laser repetition frequency of 40 MHz.

High laser trigger frequencies on the order of a few tens of MHz are needed to avoid parasitic effects from the photogenerated holes via the photoconductive gain effect with both build-up and recombination times on the millisecond timescale, as discussed in detail in references [1,2]. Depending on the laser intensity and thus on the number of photogenerated holes accumulated around the QPC, this effect shifts the quantization steps with respect to the topgate voltage, as can be seen in Supplementary Figs. S2(a) and (b). The shift of the pinch-off voltage as a function of the laser power is depicted in Fig. S2(c) for various experimental conditions. The hole accumulation happens on a ms timescale [2]. By this, the shift of the bands by the photoconductive gain effect can be regarded as quasi-static in all experiments at high repetition frequencies, as presented in the main manuscript. Most importantly, for the measurements discussed in Figs. 2, 3, and 4 of the main manuscript, the gate voltage $V_G$ is chosen such that the QPC detects either at the 1st or 2nd conductance plateau throughout all $P_{laser}$ and positions of the laser, respectively.

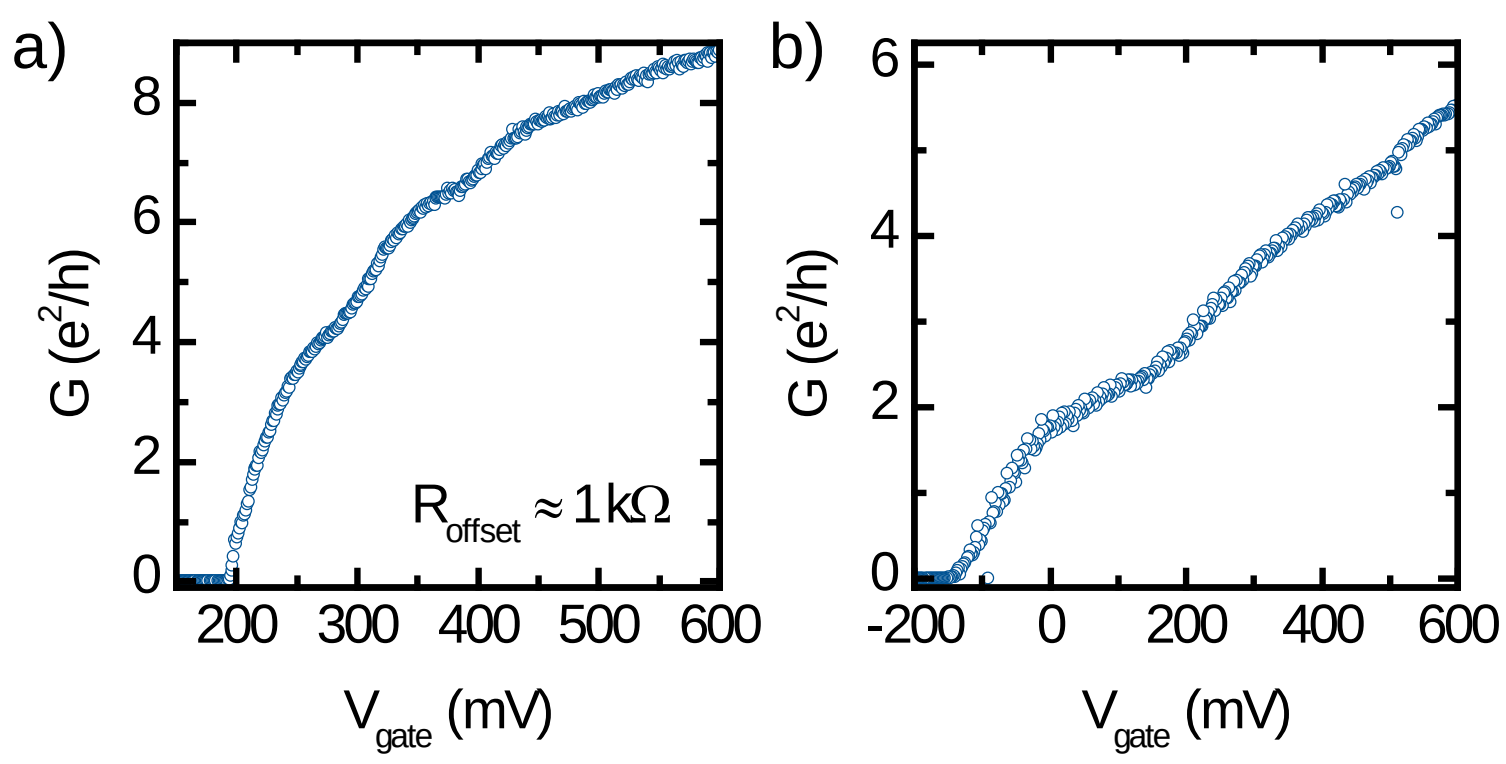

SUPPLEMENTARY FIG. S1. Conductance of the QPC used for all optoelectronic experiments measured illumination (a) and during illumination (b). In (a) a series resistance $R_{\mathrm{series}}$ is subtracted, such that the conductance plateaus occur at integer multiples of $2e^2/h$. In (b) the conductance was measured during illumination with the laser beam. In this case the series resistance is negligible.

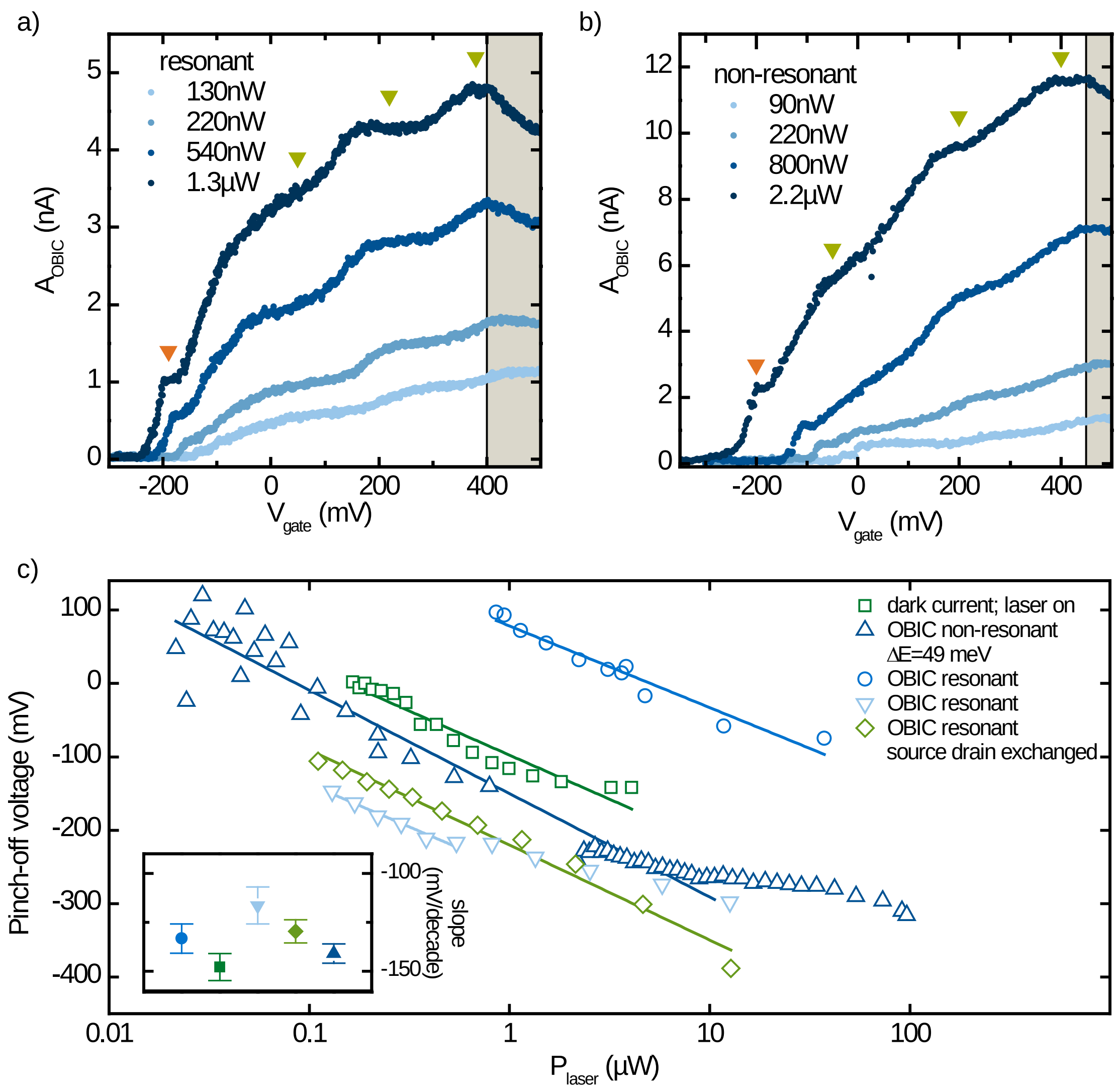

SUPPLEMENTARY FIG. S2. Quantized photocurrent for different values of the excitation power in the case of (a) resonant and (b) non-resonant excitation with an excess energy of 49 meV. (c) Pinch off voltage of the one-dimensional channel as a function of the excitation power for various measurements both resonant and non-resonant. The voltage necessary to open the channel decreases exponentially with increasing laser power due to the capacitive coupling of the accumulated photogenerated holes. The inset shows the logarithmic slope of the shift (solid lines denote exponential fits).

## 2. 2D fitting procedure

To analyze the flow patterns of the photogenerated charge carriers, we fit the distributions such as in Fig. 2(a) of the main manuscript with a two-dimensional function. The function comprises the following parts. First, we consider a Gaussian distribution centered on a circle with radius $r_{\mathrm{eff}}$ around a midpoint ($x_c$, $y_c$). This function has a width $\sigma_g$ perpendicular to the tangent of the circle and it is depicted in Fig S3(a).

$$f := e^{-\frac{\left(\sqrt{(x-x_c)^2+(y-y_c)^2}-r_{\mathrm{eff}}\right)^2}{2\sigma_g^2}} \qquad \text{(I).}$$

To account for widening of the distribution along the trajectory $r_{\mathrm{eff}} \cdot \phi$ with $r_{\mathrm{eff}}$ and $\phi$ the polar coordinates, we allow $\sigma_g$ to vary according to

$$\sigma_g = \sigma_0 \cdot \left[1 + \left(\frac{r_{\mathrm{eff}} \cdot \Phi}{a}\right)^b\right] \qquad \text{(II),}$$

where $a$ and $b$ serve as fitting parameters indicating the dependence of the width on the distance from the QPC $a$ and the widening itself $b$.

The second feature of the signal is an exponential decay of the OBIC amplitude, as already seen by Hof et al. in ref. [3]. For the present magnetic focusing data, the decay needs to be considered along the bent trajectory starting at the onset of the opaque topgate, influenced by the spatial width of the Gaussian laser spot. This is implemented by defining a convolution of an exponential decay along the orbital coordinate $r_{\mathrm{eff}} \cdot \Phi$ with decay length $l_{\mathrm{photo}}$ multiplied with a Heaviside step function starting at $r_{\mathrm{eff}} \cdot \Phi = 0$, with a Gaussian distribution (centered at $r_{\mathrm{eff}} \cdot \Phi = 0$) with a width $\sigma_c$. The latter represents the laser spot size.

$$g := \Theta(\Phi) \cdot e^{-\frac{\Phi \cdot \mathrm{R}}{l_{\mathrm{photo}}}} * \frac{1}{\sqrt{2\pi} \cdot \sigma_c} e^{-0.5 \cdot \frac{(\Phi \cdot R)^2}{\sigma_c^2}} \qquad \text{(III)}$$

The whole fitting function $h$ is a simple multiplication of $f$ and $g$.

$$h = f \cdot g \qquad \text{(IV)}$$

In supplementary Fig. S3(b) and (c) the fitting function with the exponential decay only and the additional widening can be seen.

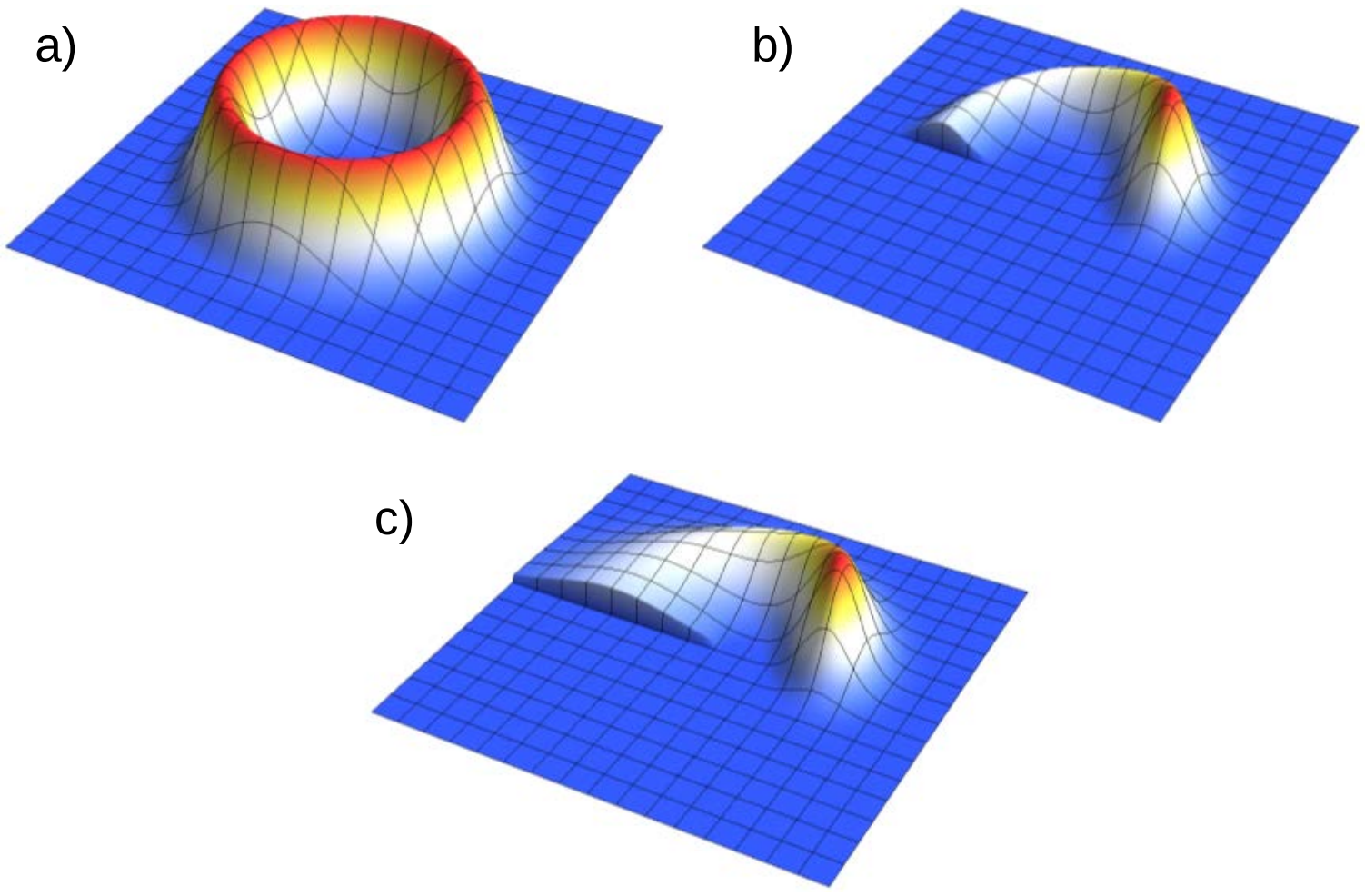

SUPPLEMENTARY FIG. S3. Step by step construction of a two-dimensional fit function for magnetic focusing data. (a) A Gaussian rotated along a circular trajectory. (b) The amplitude of the Gaussian is modulated by an exponential decay convoluted with an instrument response function. For clarity, the function is set to zero in the third quadrant. (c) The broadening of the distribution perpendicular to the circular trajectory is included.

## 3. Offset $B_0$ of the magnetic field

Generally, the applied magnetic field ***B*** at the position of the samples in the cryostat is calibrated by the help of Hall probes independently of the presented experiments. In the data in Figs. 2(d) and (e) as well as 3(a) to (d), however, an offset of the response of the QPC to the magnetic field axis is detectable. We interpret it to be caused by a slight asymmetry in the direct electrostatic environment of the source contact of the QPC or a scatterer in vicinity of QPC [5,6,7], varying for each cool-down of the sample. The two values $B_0$ = +20 mT in Figs. 3(a) and (b) and $B_0$ = -15 mT in Figs. 3(c) and (d) were taken from two different measurement periods in between which the sample was cycled to room temperature and back to liquid helium temperature. For low magnetic fields close to zero, it is difficult to tell if a certain trajectory is bent with a large radius or if it is just slightly kinked due to the effect of a possible scatterer. The underlying electrostatic environment might have been changed between the non-resonant and resonant measurements as the sample was heated to room temperature in between. By this, the two different offsets around $B_0$ = +20 mT [Fig. 3(a) and (b)] and $B_0$ = -15 mT [Fig. 3(c) and (d)] can be understood.

In principle, the wave function in QPC favors certain *k*-direction of electrons entering the QPC [4,6]. In Figs. 3(a) to (d) in the main manuscript, however, we detect very similar offsets of $B_0$ for the 1$^{st}$ and 2$^{nd}$ conductance plateaus. Therefore, we believe that the parity of the wavefunctions in the QPC plays only a minor role to explain the observed offset in magnetic fields.

## 4. Theoretical values of the slope $b$

Formula (1) in the main article states the inverse proportionality of the cyclotron radius with the electron momentum with a proportionality factor of $b = \frac{m^*|\boldsymbol{v}|}{q} = \frac{|\boldsymbol{p}|}{q}$. With the absolute value of the electron momentum $|\boldsymbol{p}| = \sqrt{2 \cdot m^* \cdot \Delta\mathrm{E}}$ , the effective electron mass of $m* = 0.067 \cdot m_0$, a maximum excess energy $\Delta E = 60\mathrm{meV}$, and the elementary charge $e$ yield an upper limit for the slope $b$ to be $213\ \mu\mathrm{m} \cdot \mathrm{mT}$. If one takes into account the energy portion transferred to the heavy holes in the valence band, one expects an electron excess energy of 49 meV, yielding a theoretical value for $b$ to be $192\ \mu\mathrm{m} \cdot \mathrm{mT}$.

## 5. Simulations for electron-electron scattering

The Monte Carlo simulations are performed as discussed in the following. Firstly, the source contact is represented as a rectangular grid [Suppl. Fig. S4]. The size of this grid is 20 µm in y-direction (-10 ; 10) and 10 µm in x-direction (0 ; 10) with a grid step size of 0.25 µm. Each grid point denotes a position vector $\vec{r} = \binom{x}{y}$. The sample design is introduced by implementing circuit boundaries with a certain reflectivity (specularity) $0 \leq p_{\mathrm{spec}} \leq 1$ with typical values around $p_{\mathrm{spec}} = 0.8$, as recently discussed in ref.[9]. The "detector-QPC" is defined as the funnel shaped end region of the circumference of the grid. If a particle reaches an x-value smaller than 4 µm, it is assumed to be collected by the QPC.

From each grid point, electrons initially start with a randomly directed velocity vector $|\vec{v}| = 0.2\,\frac{\mu\mathrm{m}}{\mathrm{ps}}$. After each time period $\Delta t = 0.2$ ps, electrons will encounter another electron and they will be scattered by a randomly chosen angle $\alpha$ uniformly distributed within the interval $-25° \leq \alpha \leq 25°$ with a probability $p$. The value for the angle distribution is estimated from Fig. 4a in reference [8]. This simple model assumes that an electron has a spatial extension of its de-Broglie-wavelength in average. For an electron density of $n$ = 2.75 x $10^{11}$ $\mathrm{cm}^{-2}$, the de-Broglie wavelength is 48 nm. Between two electron scattering processes, the trajectory is influenced by the magnetic field, and

thus, the path will always be deflected by an angle $\alpha_{\text{mag}} = \frac{|\vec{v}|\cdot\Delta t}{r_{cyclo}} \cdot \frac{360°}{2\cdot\pi}$. In the presented simulation, fields up to ±50 mT are assumed, corresponding to a minimum classical cyclotron orbit of $r_{\text{cyclo}} = 1.74\ \mu m$. The influence of the magnetic field is implemented by rotating the velocity vector after each time interval $\Delta t$ by $\alpha_{\text{mag}}$, additionally to the random rotation due to scattering. If a trajectory reaches the detector within a certain maximum number of time intervals $N \leq 2000$ (equivalent to a maximum total distance of $s_{\text{max}} = |\vec{v}| \cdot \Delta t \cdot N_{\text{max}} = 80$ µm), it is registered as a hit. Each individual trace that reaches the detector contributes to the accumulated count for its starting point with a factor $e^{\frac{s}{l_{\text{mfp}}}}$ for considering momentum scattering.

For the presented simulations, the trajectories are weighted by $l_{\text{mfp}} = 30$ µm to achieve reasonable agreement between the experimentally observed $l_{\text{photo}}$ and simulation. The experimentally observed $l_{\text{photo}}$ lies on the order of the electronic mean free path of 15.1 µm. Generally, the mean free path is a length scale measured in electron transport experiments and it is pointing in the direction of the electric field (according to Drude theory) and therefore, it is an averaged value. The length $l_{\text{mfp}}$ in the simulations, however, is a weighting factor along the actual trajectory of the photogenerated electron, and therefore, it has to be assumed to be significantly longer than the quantity measured in transport experiments. To account for non-perfectly reflecting boundaries of the circuit we additionally multiply the final value for each trajectory with a number $(p_{\text{spec}})^{N_{\text{boundary}}}$ ($N_{\text{boundary}}$: number of boundary hits).

In the simulation, from each point of the grid 100 traces are started. The simulated raw data are convoluted with a Gaussian kernel with $\sigma = 1$ µm to account for the laser broadening in the experiment. Plotting the number of hits for each point as a two-dimensional contour plot, one acquires a simulated distribution as shown in Fig. 4(c), very similar to the experimental data. To extract spatial information about the simulated distribution, we use the same fitting algorithm that we use for the experimental data (cf. Fitting procedure above).

Finally, we point out that the qualitative outcome of the simulations do not depend on the exact range of $\alpha$, as long as a “small angle scattering” is assumed. For large angle scattering, we cannot reproduce the data.

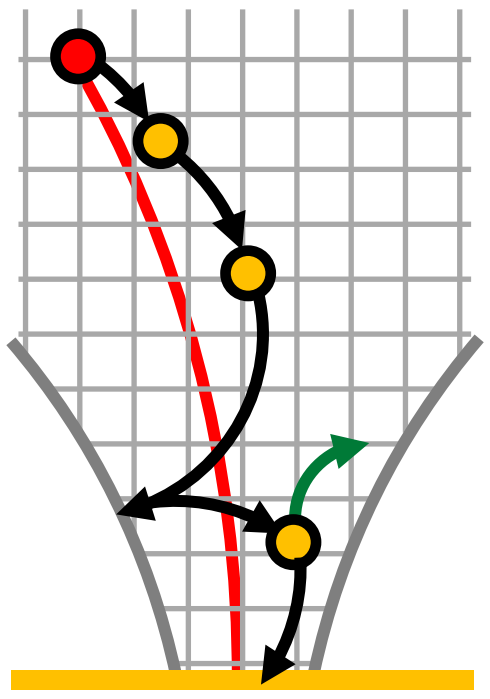

SUPPLEMENTARY FIG. S4. Schematic sketch of mechanism for increase of radius of electronic distribution due to multiple scattering events. For details see text above.

REFERENCES.